\documentclass[aps,prl,reprint,superscriptaddress]{revtex4-1}
\usepackage{graphicx}
\usepackage[ansinew]{inputenc}
\usepackage{array}
\usepackage{color}
\usepackage{amsmath}
\usepackage{amsxtra}
\usepackage{amstext}
\usepackage{amssymb}
\usepackage{latexsym}

\begin{document}

%Title of paper
\title{Real-time evolution of a laser-dressed Helium atom: Attosecond-resolved two-color photoionization study}

\author{Niranjan Shivaram}

\author{Henry Timmers}
\affiliation{Department of Physics, University of Arizona, Tucson, AZ, 85721 USA.}

\author{Xiao-Min Tong}
\affiliation{Center for Computational Sciences, University of Tsukuba, Ibaraki 305-8573, Japan.}

\author{Arvinder Sandhu}
\email[]{sandhu@physics.arizona.edu}
\affiliation{Department of Physics, University of Arizona, Tucson, AZ, 85721 USA.}

\date{\today}

\begin{abstract}
Using extreme-ultraviolet attosecond-pulse-trains, we investigate the photoionization dynamics of a Helium atom in the presence of moderately-strong ($\sim10^{12}$Wcm$^{-2}$) femtosecond laser pulses. The electronic structure of a laser-dressed atom is traced in real-time through precision measurements of ion-yields and photo-electron angular distributions. Quantum interferences between photo-excitation paths are interpreted using the Floquet formalism. As the laser pulse intensity ramps on femtosecond timescales, we observe transitions between ionization channels mediated by different atomic resonances. The quantum phase of interfering paths is extracted for each channel and compared with simulations. Our results elucidate photoionization mechanisms in strong-fields and open the doors for photo-absorption/ionization control schemes.
\end{abstract}

% insert suggested PACS numbers in braces on next line
\pacs{}
% insert suggested keywords - APS authors don't need to do this
%\keywords{}

%\maketitle must follow title, authors, abstract, \pacs, and \keywords
\maketitle

The recent advances in `attosecond science' have given a new impetus to the study of atomic and molecular phenomena by providing direct real-time access to electron dynamics\cite{krausz2009}. Experiments in this regime are typically conducted using extreme-ultraviolet (XUV) attosecond pulses or pulse trains along with precisely synchronized strong-field femtosecond near-infrared (IR) laser pulses, to obtain new insights into dynamics of electronically excited systems\cite{krausz2009,sandhu2008,wang2010}. As the roots of `attosecond science' lie in the strong-field concepts developed in 1990's \cite{corkum1993}, the application of new attosecond techniques to refine our understanding of atomic/molecular dynamics in strong fields is of particular interest\cite{krausz2009}.

Here, we report precision real-time measurements of the transient non-equilibrium electronic structure of Helium in intense fields. We investigate the quantum interferences in two-color photo-ionization pathways using XUV attosecond pulse trains (APT) and variable strength near-infrared (IR) laser fields. As the field intensity changes on femtosecond timescales, we observe switching between ionization channels. We find that yield from each resonance-mediated ionization channel oscillates with a specific phase. We interpret this quantum phase using Floquet interaction model. Numerical calculations using time-dependent Schr{\"o}dinger equation (TDSE) serve to elucidate the important role of Floquet interferences in photo-excitation and ionization.

\begin{figure}[t]
\includegraphics[width=0.49 \textwidth]{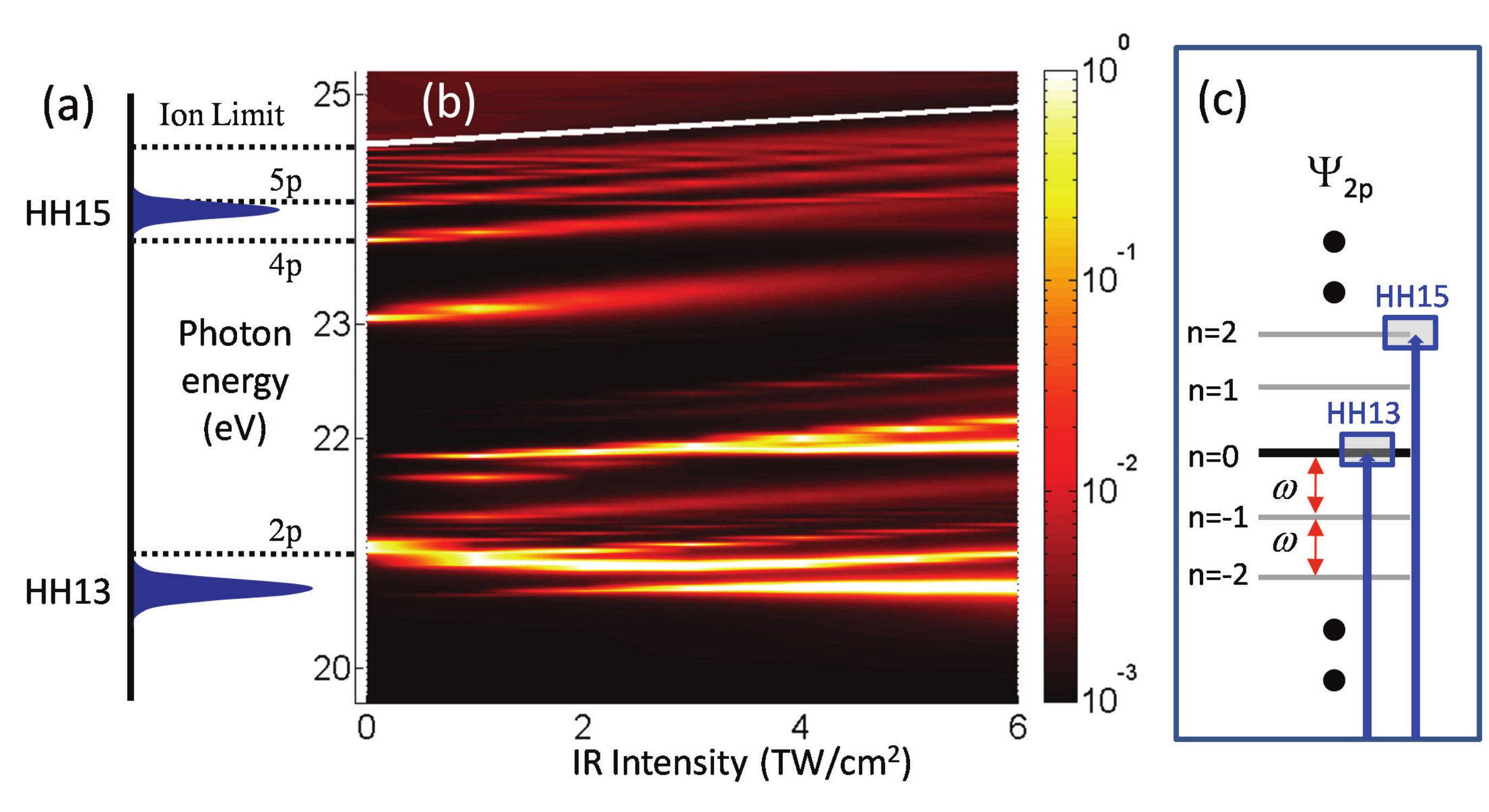}
\caption{\label{fig:1} (a) XUV spectrum and the relevant He states (b) Calculated XUV photoabsorption cross-section as a function of laser intensity. (c) Floquet manifold showing one-photon spaced components of IR-dressed 2p state and the two interfering paths associated with HH13 and HH15 excitation.}
\end{figure}

We use amplified 65fs, 785nm IR pulses of 1.5mJ energy, which are split into two parts. One part is focused onto a Xe gas-filled hollow waveguide to generate APT of high-harmonics(HH). The APT along with the co-propagating driving infrared pulse (IR$_{d}$) is focused onto a He gas jet using a toroidal mirror. The second part, a probe pulse (IR$_{p}$), goes to a delay stage and is focused on the He target with a 50cm lens. The schematic of our experimental set-up is provided in \cite{shivaram2010}. Photo-electrons are imaged using a velocity-map-imaging (VMI) setup. He$^{+}$ ions are spatially imaged such that there is a one-to-one correspondence between the point of origin and the point they hit the detector, allowing us to eliminate Gouy phase averaging and obtain a high-quality signal\cite{shivaram2010}.

Fig. \ref{fig:1}(a) schematically shows the experimental HH spectra relative the unperturbed He resonances. The 15th harmonic is resonant with the 5p electronic state and 13th harmonic is slightly below 2p resonance. Two other harmonics that we observe, i.e. 11th and 17th, are much weaker (20 times lower) and non-resonant, hence they do not play a significant role in this study. In Fig. \ref{fig:1}(b) we show the photo-absorption cross-section of He as a function of photon energy and peak IR intensity calculated using the method described in \cite{tong2010stru}. Clearly, the discrete atomic resonances evolve into a complicated structure even in a moderately intense laser field of the order of 10$^{12}$Wcm$^{-2}$. The higher excited-states (3p, 4p, 5p, etc.) exhibit positive shifts, which can be approximated by the ponderomotive effect. The low-lying 2p state exhibits a negative shift and develops multiple branches. At intensities around 5$\times$10$^{12}$Wcm$^{-2}$, the excited state structure bears very little resemblance to the unperturbed case.

In previous experiments with APT and IR, it has been observed that the He$^{+}$ ion yield oscillates as a function of time delay with half-IR-cycle periodicity \cite{johnsson2007,ranitovic2010,tong2010mech}. This has been interpreted as interference between wave packets generated by successive bursts in the APT\cite{johnsson2007,riviere2009,holler2011}. Using Floquet interaction picture\cite{chu2004}, it has been shown that the interference between different Fourier components of an IR dressed state leads to this oscillatory variation of ionization probability at $2\omega$ frequency\cite{tong2010mech,tong2010floq}. Importantly, recent theoretical work\cite{tong2010floq} has raised crucial questions about the role of different ionization paths and the phase of ionization signal that have not been addressed by the experimental studies conducted so far. Here we measure the $\it{phase}$ of the ion-yield oscillations and investigate its variation in terms of the evolution of laser-dressed atomic structure. We demonstrate that for a given IR-dressed atomic resonance (Fig. \ref{fig:1}(c)), the quantum phase difference between the dipole transitions to different Fourier components determines the oscillation phase of the ionization signal.

In our XUV+IR ionization measurements, the IR field results from a combination of two IR pulses. The weaker pulse (IR$_{d}$) is phase-locked to the APT and the stronger pulse (IR$_{p}$) is time delayed relative to the APT\cite{shivaram2010}. For two main harmonics (HH13 and HH15 in our case), the probability of ionization by the XUV and combined IR field can be obtained for a given Floquet state as \cite{tong2010floq}
\begin{equation}
P(\tau) \propto \left|M_0 f_0+M_2 f_2 e^{-i(2\omega\tau + 2\delta_0 -2\delta(\tau) +\phi)}\right|^2
\end{equation}
where $M_0(A(\tau))$ and  $M_2(A(\tau))$ are the IR-intensity dependent  matrix elements representing transitions to the direct and two-photon-dressed Fourier component. The terms $f_0$ and $f_2$ are the strengths of two harmonics, $\omega$  is the central frequency of the IR, $\tau$  is the time delay between XUV and probe IR, $\delta_0$  is the phase at which the attosecond pulse is locked to the driver IR field.  The quantum phase difference between $M_0$ and $M_2$, which is an important quantity in the paper, is represented by $\phi$. Figure \ref{fig:1}(c) diagramatically shows the Floquet manifold associated with 2p resonance and the two interfering excitation paths discussed above.

The delay dependence of combined IR amplitude and phase, i.e. $A(\tau)$ and $\delta(\tau)$, is given as $A(\tau) = {(A_{p}^{2}+A_{d}^{2}+2A_{p}A_{d}\cos(\omega\tau))}^{1/2}$
and 
$\delta(\tau) = \sin^{-1}(A_d\sin(\omega\tau)/A(\tau))$
, where $A_p$ and $A_d$ are the amplitudes of the probe and driver IR fields. For a weak driver-IR field ($A_{d} \ll A_{p}$), the dominant frequencies in $P(\tau)$ are $1\omega$  and $2\omega$. We define the normalized amplitude of ion-yield oscillation as $P_{osc}(\tau)=(P(\tau)-P_{avg})/P_{avg}$, where $P_{avg}$ is the one-cycle average. The $P_{osc}$ in the weak driver case can then be approximated as
\begin{equation}
P_{osc}(\tau) = T_{1}\cos(\omega\tau)+T_{2}\cos(2\omega\tau+2\delta_{0}+\phi)
\end{equation}
 where $T_{1}$ , $T_{2}$ are the amplitude factors. 

\begin{figure}[t]
\includegraphics[width=0.49 \textwidth]{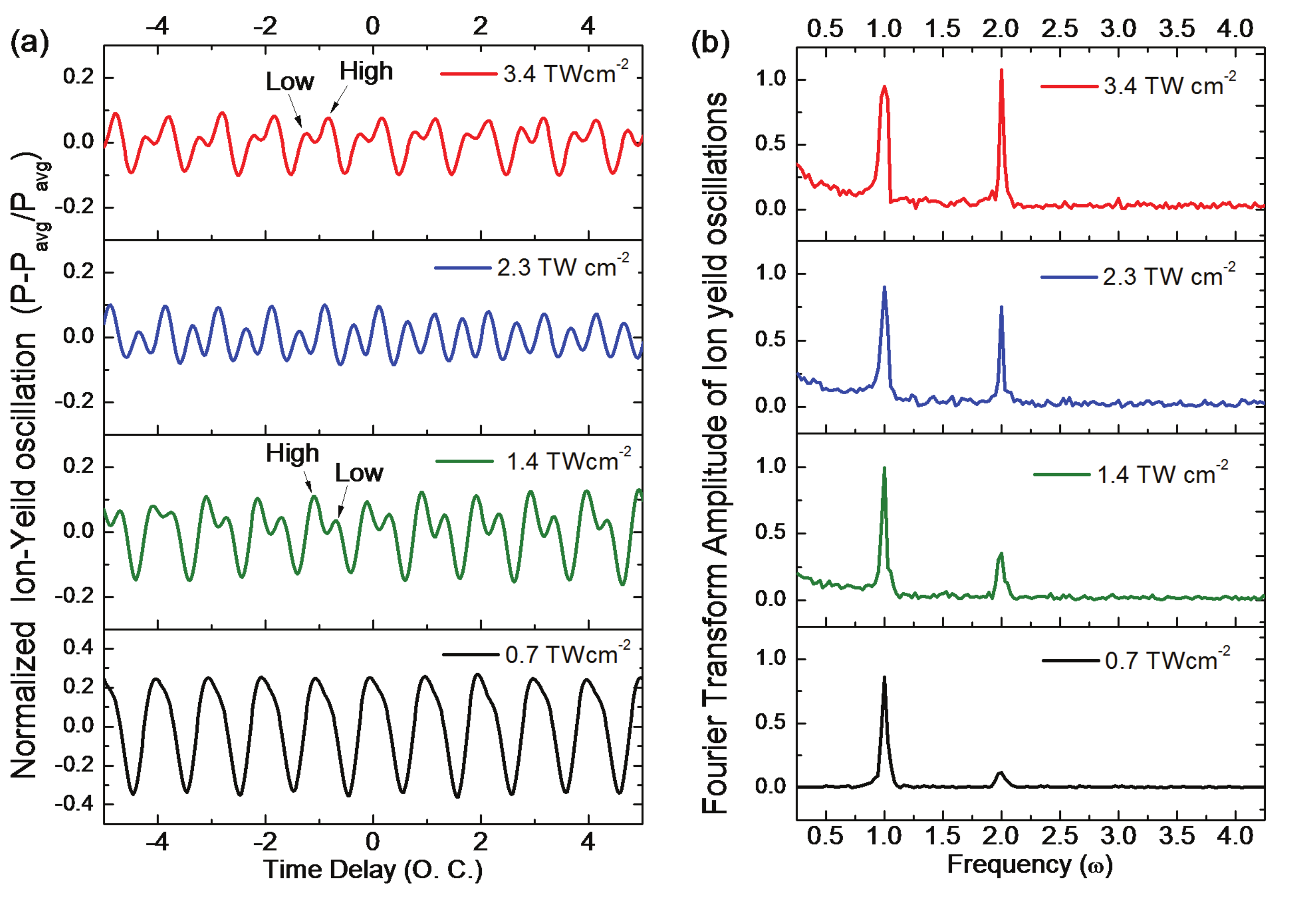}
\caption{\label{fig:2}(a) Normalized He$^+$ ion-yield oscillations for XUV and 2IR pulses (driver and probe) at different probe peak intensities (o.c. denotes optical cycles). Asymmetry in the double peak oscillation structure is reversed between 1.4 TWcm$^{-2}$ and 3.4 TWcm$^{-2}$. (b) Fourier amplitude of ion-yield oscillations shows two prominent frequencies.}
\end{figure}

Figure \ref{fig:2}(a) shows normalized experimental He$^+$ ion-yield as a function of the time delay between APT+IR$_d$ and probe IR$_p$ at different probe IR intensities. The driver IR intensity is less than 10$^{10}$Wcm$^{-2}$. We observe a distinct oscillation structure with one-cycle and half-cycle components in accordance with equation (2). The oscillation at $1\omega$ frequency, which arises from the intensity modulation due to the IR-IR interference, acts as a reference with respect to which we can robustly measure the phase $\phi$ of the Floquet path interferences occurring at $2\omega$. The 2IR method thus allows use to compensate for the interferometric drifts and other experimental variations. Furthermore, this approach also provides a method to measure $\delta_0$, which represents the timing of attosecond bursts relative to the peak of driving IR field.

Fig. \ref{fig:2}(b) shows the Fourier transform of ion-yield oscillations over the time-delay range that spans more than 20 optical cycles. At low intensities, the $1\omega$ component dominates, however, as the probe intensity is increased, the $2\omega$ component increases. This is expected as the IR amplitude modulation decreases and the two-photon dressed Floquet contribution (i.e. $M_2$ in equation (1)) becomes increasingly important with intensity, leading to a stronger interference signal. Importantly, the oscillatory structure in Fig. \ref{fig:2}(a) at two probe intensities, namely, 1.4 TW cm$^{-2}$ and 3.4 TW cm$^{-2}$ is very different. The asymmetric double-peak structure at 1.4 TW cm$^{-2}$ shows the left-peak to be higher, whereas, at 3.4 TW cm$^{-2}$ the situation is reversed and the right-peak is higher. This difference in oscillation structure is a direct manifestation of the change in phase relationship between the $2\omega$ and $1\omega$ components. The intensity dependent change in relative phase originates from the change in value of $\phi$ in eq.(1), which represents the phase difference between the two interfering contributions. Before we quantitatively discuss this phase, it is important to identify the Floquet states contributing to the ionization.

\begin{figure}[t]
\includegraphics[width=0.49 \textwidth]{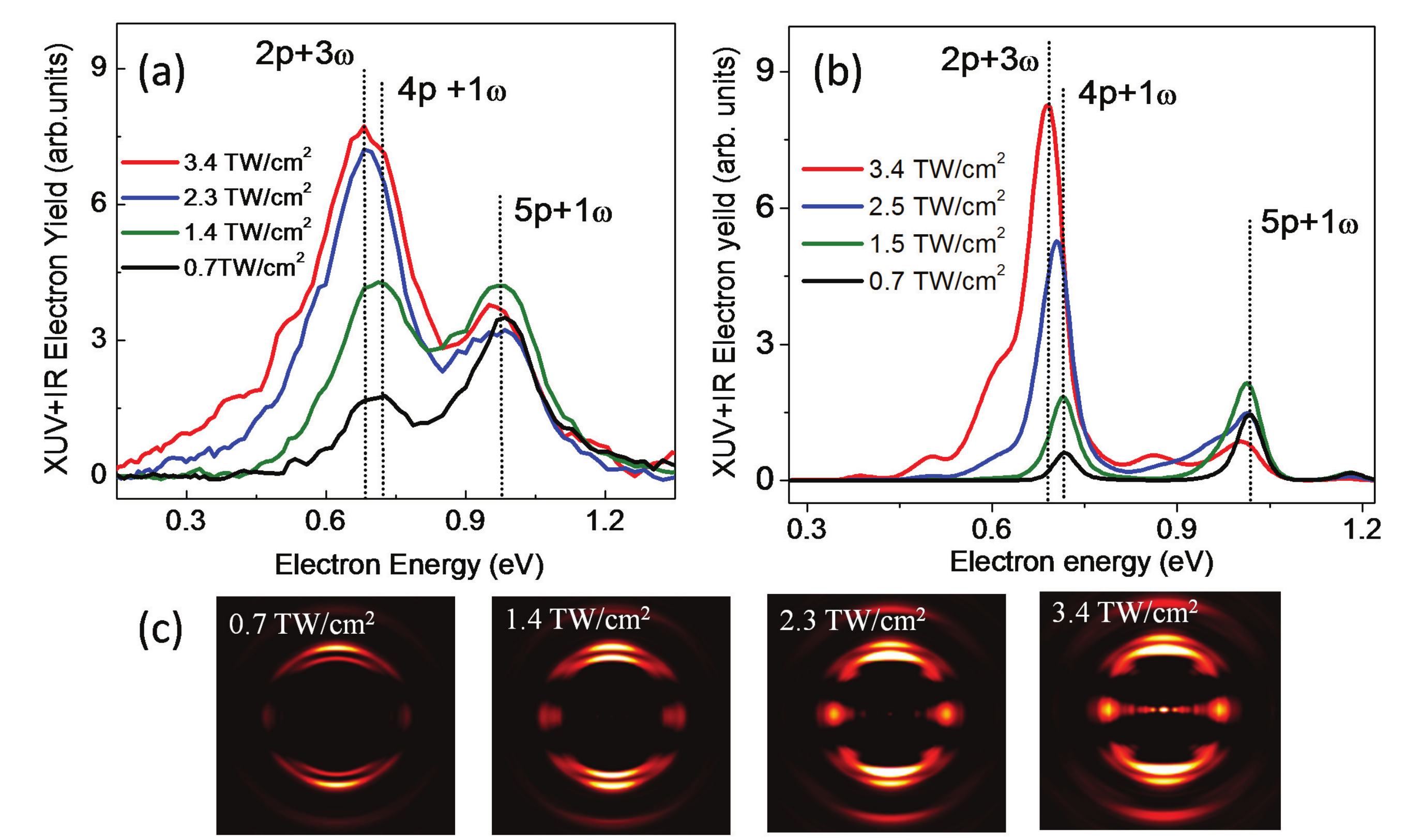}
\caption{\label{fig:3} (a) Experimental XUV+IR photo-electron spectrum of He at different IR probe intensities. As the IR intensity is increased the $2p+3\omega$ peak becomes dominant. (b) Calculated photo-electron spectrum (c) Experimental VMI of photo-electrons at different IR probe  intensities.}
\end{figure}

To identify the Floquet paths, we utilize photoelectron spectroscopy. Fig. \ref{fig:3}(a) shows the experimental electron spectra at probe intensities used in Fig. \ref{fig:2}. The observed electron peaks in Fig. \ref{fig:3}(a) are associated with IR ionization of XUV excited 5p, 4p and 2p atomic states. At low intensities, the ionization is mediated by $5p$ resonance, and we observe a strong peak corresponding to $5p+1\omega$ process. This is expected as 15th harmonic is initially resonant with 5p state. As the intensity is increased, the $4p+1\omega$ starts contributing. At higher intensities the $2p+3\omega$ channel dominates the ionization signal. This observation is in accord with Fig. \ref{fig:1}(b) as the 2p structure Stark shifts downward in energy with increasing intensity, becoming resonant at higher intensities. 

TDSE calculations for comparable intensity parameters also yield similar results (Fig. \ref{fig:3}(b)). The angle-resolved photoelectron images in Fig. \ref{fig:3}(c) also show that as intensity is increased towards 3.4 TWcm$^{-2}$, side lobes corresponding to the `g-wave' structure appear. This another indication of 3-photon ionization of XUV excited state 2p state. Thus, the results from Fig. \ref{fig:3} confirm that the dominant two-color ionization pathway changes from the 5p-mediated ionization at low intensities to  2p-mediated ionization signal at higher intensities. Next, we extract the phases for various resonance mediated ionization channels and establish quantitative relationship between the strong-field variation of atomic structure and intensity dependent phase change observed in ion-yield oscillations of figure \ref{fig:2}.

\begin{figure}[t]
\includegraphics[width=0.49 \textwidth]{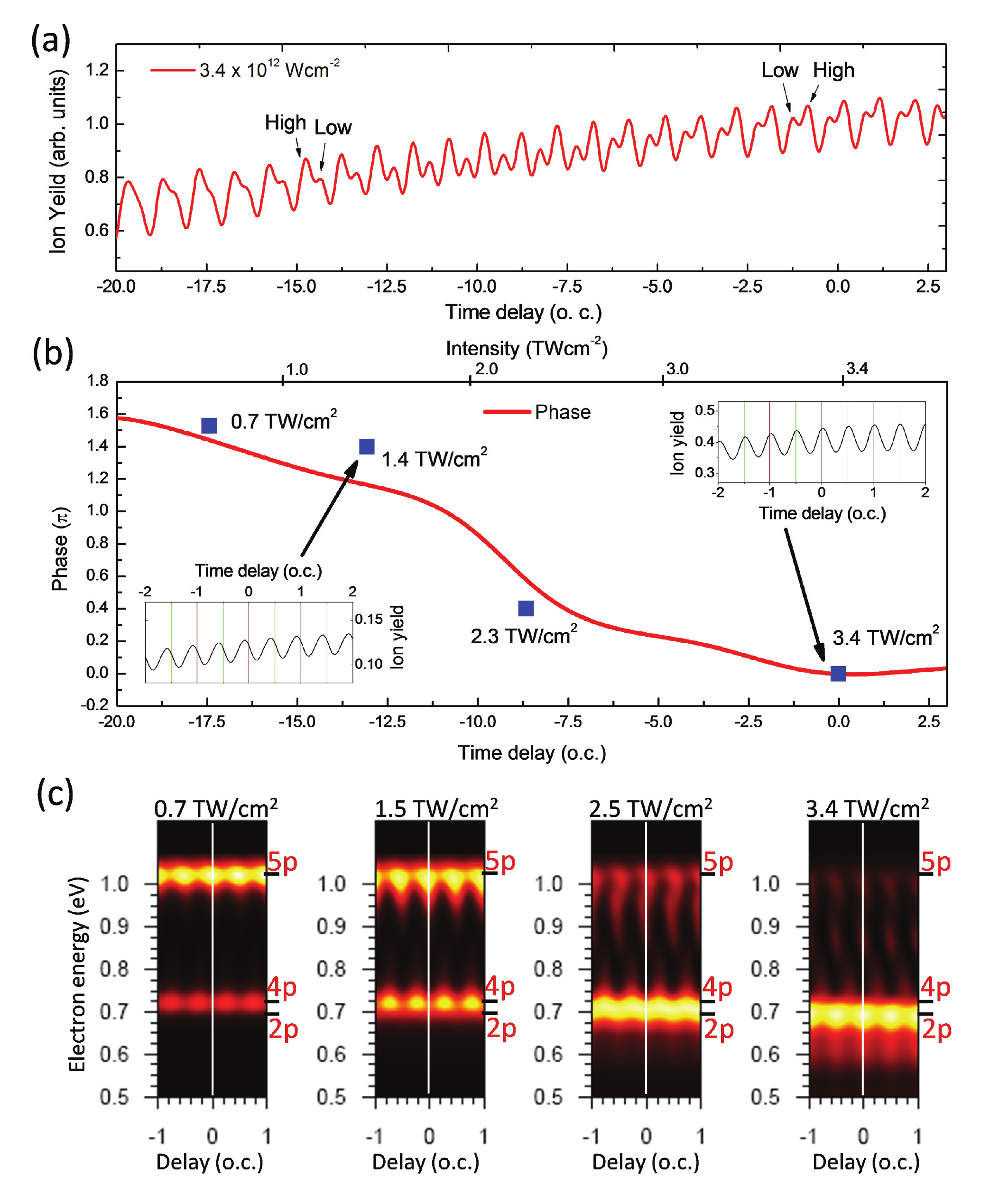}
\caption{\label{fig:4} He$^{+}$ ion yield as a function of time-delay at 3.4 TWcm$^{-2}$ probe intensity. (b) The phase of the $2\omega$ component of ion-yield oscillations with time delay. (c) TDSE results showing the energy-resolved oscillations in electron yield of various resonance mediated ionization channels.}
\end{figure}

Figs. \ref{fig:2} and \ref{fig:3} show that even moderately intense laser pulses can significantly modify an atom and its ionization dynamics. Now we probe these transient dynamics in real time. Fig. \ref{fig:4}(a) shows the raw He$^+$ ion yield as a function of time-delay at 3.4 TW cm$^{-2}$ probe intensity. The negative time-delay axis implies that the IR probe arrives ahead of the XUV pulse. As the time-delay changes from -20 o.c. towards zero, the asymmetric $2\omega$ oscillation structure develops and evolves from high-left-peak  asymmetry to high-right-peak asymmetry. Note that this behavior is similar to Fig. \ref{fig:2}, except that Fig. \ref{fig:2} shows the dependence of ionization signal on peak intensity, whereas, Fig. \ref{fig:4}(a) explores the dependence on intensity variation within the IR pulse profile. 

Next, we extract the phase of $2\omega$ component of ion-yield oscillations relative to the $1\omega$ component using standard Fourier-transform methods. Figure \ref{fig:4}(b) plots this phase as a function of the time-delay (solid line). We repeat the phase extraction excercise on the ionization signals in Fig. \ref{fig:2}(a) and obtain the phase of $2\omega$ component at different peak intensities. Assuming a Gaussian profile, we calibrate the time-delay axis such that each delay value corresponds to a specific instantaneous intensity (top-axis of Fig. \ref{fig:4}(b)). Thus, we can plot the phases extracted from Fig. \ref{fig:2}(a) as square dots in Fig. \ref{fig:4}(b) and compare them with delay-dependent phase curve. A good match between the two independent results demonstrates the soundness of our experimental method.

The insets in figure \ref{fig:4}(b) show the TDSE results for XUV+IR ionization oscillation at 3.4 TWcm$^{-2}$ and 1.5 TWcm$^{-2}$ with arrows pointing to comparable experimental points. At 3.4 TW cm$^{-2}$, where 2p contribution strongly dominates, the phase of $2\omega$ oscillations is zero (right-inset), implying ionization yield maximizes when APT arrives at the peak of IR-field. We  can use this fact to extract the $\delta_0$ in eq. (2) and remove APT timing offset from our phase deduction, thereby setting the starting point of experimental phase value to zero. With this adjustment, the modified form of eq. (1) becomes 
\begin{equation}
P(\tau) \propto \left|M_0 f_0+M_2 f_2 e^{-i(2\omega\tau + \phi)}\right|^2
\end{equation}
Starting from zero, as we move to longer time-delays (i.e. lower intensities), the experimental phase value increases. The phase goes through $\pi$ and reaches $1.5\pi$ at 1.4 TWcm$^{-2}$ intensity. The TDSE calculation (left-inset) also yields a similar result as the $2\omega$ oscillations in shifted in phase by $\sim 1.5\pi$ at 1.5 TWcm$^{-2}$.

The variation of phase figure \ref{fig:4}(b) can now be understood in terms of the quantum mechanical phases associated with various ionization channels and the relative dominance of different channels. The calculated energy-resolved photoelectron data in figure \ref{fig:4}(c) elucidates the role of different channels. At high intensity (3.4 TW cm$^{-2}$), the 2p resonance mediated ionization channel at 0.69 eV dominates. The oscillation of this channel has zero phase and the ionization yield peaks at zero-delay where XUV pulses arrive at the peak of IR field. For this to happen the quantum phase $\phi$ between the $M_{0}$ and $M_{2}$ transition matrix elements in eq. (3) has to be zero. In other words, 13HH and 15HH induced transitions to the Floquet components associated with 2p excitation are `in-phase' (Fig. \ref{fig:1}(c)). 

As intensity decreases, the 4p channel at 0.72 eV starts contributing substantially. Interestingly, the oscillations in this channel are completely out of phase with the 2p contribution (Fig. \ref{fig:4}(c),1.5 TW cm$^{-2}$ ). Using eq. (3), this implies that $\phi=\pi$ for 4p mediated ionization channel and the ionization yield peaks at the zeros of the IR field. In Floquet picture, the 13th and 15th HH transition matrix elements to the components of 4p Floquet state have opposite signs. In general, as the transition matrix elements are real, ideally there are only two possibilities; either $M_{0}$ and $M_{2}$ can have the same sign ($\phi=0,\pm2\pi,..$) or opposite signs ($\phi=\pm\pi, \pm3\pi,..$) and we see both varieties in Fig. \ref{fig:4}(c). At very low intensity, the 5p ionization starts dominating and oscillates with same phase as 2p (i.e. $0$,$\pm2\pi$). Thus, in our experiment, as the intensity decreases or time-delay get longer, we observe a change in phase of ion-yield oscillation, going from zero when 2p is dominant, through $\pi$ where 4p contributes substantially, and eventually towards $2\pi$ where 5p dominates.

Our work represents a direct measurement of quantum phases associated with XUV+IR ionization channels, which are not known $\it{a}$ $\it{priori}$. These observations are general and should be valid whenever XUV excitation occurs in a strong field. As the modification of electronic states depends on the instantaneous value of the IR laser intensity at the time of XUV excitation, tuning IR intensity can change contribution of the different matrix elements to the interfering terms. By precisely controlling the XUV spectrum, IR intensity and the time-delay between the APT and IR, it is possible to control the ionization dynamics. A recent experiment\cite{ranitovic2011} demonstrating XUV transparency validates some of these ideas.

In conclusion, the two-color photo-ionization of He by XUV and IR pulses  represents a relatively simple yet rich system for exploration of light-matter interaction. We show that even a moderately strong field, often used in XUV-IR pump-probe studies, can substantially modify electronic structure and ionization dynamics on attosecond timescales. We identify different ionization channels and measure the quantum phases associated with these Floquet paths, which is the first such measurement to our knowledge. Understanding the evolution of atomic structure in strong fields and its interaction with attosecond XUV pulses can provide us with a knob for fine control of photo-dynamics.

%\begin{acknowledgments}
This work was supported by the National Science Foundation (NSF) under contract PHY-0955274.
%\end{acknowledgments}

% Create the reference section using BibTeX:
\bibliography{RefsShivaram}

%merlin.mbs apsrev4-1.bst 2010-07-25 4.21a (PWD, AO, DPC) hacked
%Control: key (0)
%Control: author (8) initials jnrlst
%Control: editor formatted (1) identically to author
%Control: production of article title (-1) disabled
%Control: page (0) single
%Control: year (1) truncated
%Control: production of eprint (0) enabled
\begin{thebibliography}{14}%
\makeatletter
\providecommand \@ifxundefined [1]{%
 \@ifx{#1\undefined}
}%
\providecommand \@ifnum [1]{%
 \ifnum #1\expandafter \@firstoftwo
 \else \expandafter \@secondoftwo
 \fi
}%
\providecommand \@ifx [1]{%
 \ifx #1\expandafter \@firstoftwo
 \else \expandafter \@secondoftwo
 \fi
}%
\providecommand \natexlab [1]{#1}%
\providecommand \enquote  [1]{``#1''}%
\providecommand \bibnamefont  [1]{#1}%
\providecommand \bibfnamefont [1]{#1}%
\providecommand \citenamefont [1]{#1}%
\providecommand \href@noop [0]{\@secondoftwo}%
\providecommand \href [0]{\begingroup \@sanitize@url \@href}%
\providecommand \@href[1]{\@@startlink{#1}\@@href}%
\providecommand \@@href[1]{\endgroup#1\@@endlink}%
\providecommand \@sanitize@url [0]{\catcode `\\12\catcode `\$12\catcode
  `\&12\catcode `\#12\catcode `\^12\catcode `\_12\catcode `\%12\relax}%
\providecommand \@@startlink[1]{}%
\providecommand \@@endlink[0]{}%
\providecommand \url  [0]{\begingroup\@sanitize@url \@url }%
\providecommand \@url [1]{\endgroup\@href {#1}{\urlprefix }}%
\providecommand \urlprefix  [0]{URL }%
\providecommand \Eprint [0]{\href }%
\providecommand \doibase [0]{http://dx.doi.org/}%
\providecommand \selectlanguage [0]{\@gobble}%
\providecommand \bibinfo  [0]{\@secondoftwo}%
\providecommand \bibfield  [0]{\@secondoftwo}%
\providecommand \translation [1]{[#1]}%
\providecommand \BibitemOpen [0]{}%
\providecommand \bibitemStop [0]{}%
\providecommand \bibitemNoStop [0]{.\EOS\space}%
\providecommand \EOS [0]{\spacefactor3000\relax}%
\providecommand \BibitemShut  [1]{\csname bibitem#1\endcsname}%
\let\auto@bib@innerbib\@empty
%</preamble>
\bibitem [{\citenamefont {Krausz}\ and\ \citenamefont
  {Ivanov}(2009)}]{krausz2009}%
  \BibitemOpen
  \bibfield  {author} {\bibinfo {author} {\bibfnamefont {F.}~\bibnamefont
  {Krausz}}\ and\ \bibinfo {author} {\bibfnamefont {M.}~\bibnamefont
  {Ivanov}},\ }\href@noop {} {\bibfield  {journal} {\bibinfo  {journal} {Rev.
  Mod. Phys.}\ }\textbf {\bibinfo {volume} {81}},\ \bibinfo {pages} {163}
  (\bibinfo {year} {2009})}\BibitemShut {NoStop}%
\bibitem [{\citenamefont {Sandhu}\ \emph {et~al.}(2008)\citenamefont {Sandhu},
  \citenamefont {Gagnon}, \citenamefont {Santra}, \citenamefont {Sharma},
  \citenamefont {Li}, \citenamefont {Ho}, \citenamefont {Ranitovic},
  \citenamefont {Cocke}, \citenamefont {Murnane},\ and\ \citenamefont
  {Kapteyn}}]{sandhu2008}%
  \BibitemOpen
  \bibfield  {author} {\bibinfo {author} {\bibfnamefont {A.~S.}\ \bibnamefont
  {Sandhu}}, \bibinfo {author} {\bibfnamefont {E.}~\bibnamefont {Gagnon}},
  \bibinfo {author} {\bibfnamefont {R.}~\bibnamefont {Santra}}, \bibinfo
  {author} {\bibfnamefont {V.}~\bibnamefont {Sharma}}, \bibinfo {author}
  {\bibfnamefont {W.}~\bibnamefont {Li}}, \bibinfo {author} {\bibfnamefont
  {P.}~\bibnamefont {Ho}}, \bibinfo {author} {\bibfnamefont {P.}~\bibnamefont
  {Ranitovic}}, \bibinfo {author} {\bibfnamefont {C.~L.}\ \bibnamefont
  {Cocke}}, \bibinfo {author} {\bibfnamefont {M.~M.}\ \bibnamefont {Murnane}},
  \ and\ \bibinfo {author} {\bibfnamefont {H.~C.}\ \bibnamefont {Kapteyn}},\
  }\href@noop {} {\bibfield  {journal} {\bibinfo  {journal} {Science}\ }\textbf
  {\bibinfo {volume} {322}},\ \bibinfo {pages} {1081} (\bibinfo {year}
  {2008})}\BibitemShut {NoStop}%
\bibitem [{\citenamefont {Wang}\ \emph {et~al.}(2010)\citenamefont {Wang},
  \citenamefont {Chini}, \citenamefont {Chen}, \citenamefont {Zhang},
  \citenamefont {He}, \citenamefont {Cheng}, \citenamefont {Wu}, \citenamefont
  {Thumm},\ and\ \citenamefont {Chang}}]{wang2010}%
  \BibitemOpen
  \bibfield  {author} {\bibinfo {author} {\bibfnamefont {H.}~\bibnamefont
  {Wang}}, \bibinfo {author} {\bibfnamefont {M.}~\bibnamefont {Chini}},
  \bibinfo {author} {\bibfnamefont {S.~Y.}\ \bibnamefont {Chen}}, \bibinfo
  {author} {\bibfnamefont {C.~H.}\ \bibnamefont {Zhang}}, \bibinfo {author}
  {\bibfnamefont {F.}~\bibnamefont {He}}, \bibinfo {author} {\bibfnamefont
  {Y.}~\bibnamefont {Cheng}}, \bibinfo {author} {\bibfnamefont
  {Y.}~\bibnamefont {Wu}}, \bibinfo {author} {\bibfnamefont {U.}~\bibnamefont
  {Thumm}}, \ and\ \bibinfo {author} {\bibfnamefont {Z.~H.}\ \bibnamefont
  {Chang}},\ }\href@noop {} {\bibfield  {journal} {\bibinfo  {journal}
  {Physical Review Letters}\ }\textbf {\bibinfo {volume} {105}},\ \bibinfo
  {pages} {143002} (\bibinfo {year} {2010})}\BibitemShut {NoStop}%
\bibitem [{\citenamefont {Corkum}(1993)}]{corkum1993}%
  \BibitemOpen
  \bibfield  {author} {\bibinfo {author} {\bibfnamefont {P.~B.}\ \bibnamefont
  {Corkum}},\ }\href@noop {} {\bibfield  {journal} {\bibinfo  {journal} {Phys.
  Rev. Lett.}\ }\textbf {\bibinfo {volume} {71}},\ \bibinfo {pages} {1994}
  (\bibinfo {year} {1993})}\BibitemShut {NoStop}%
\bibitem [{\citenamefont {Shivaram}\ \emph {et~al.}(2010)\citenamefont
  {Shivaram}, \citenamefont {Roberts}, \citenamefont {Xu},\ and\ \citenamefont
  {Sandhu}}]{shivaram2010}%
  \BibitemOpen
  \bibfield  {author} {\bibinfo {author} {\bibfnamefont {N.}~\bibnamefont
  {Shivaram}}, \bibinfo {author} {\bibfnamefont {A.}~\bibnamefont {Roberts}},
  \bibinfo {author} {\bibfnamefont {L.}~\bibnamefont {Xu}}, \ and\ \bibinfo
  {author} {\bibfnamefont {A.}~\bibnamefont {Sandhu}},\ }\href@noop {}
  {\bibfield  {journal} {\bibinfo  {journal} {Optics Letters}\ }\textbf
  {\bibinfo {volume} {35}},\ \bibinfo {pages} {3312} (\bibinfo {year}
  {2010})}\BibitemShut {NoStop}%
\bibitem [{\citenamefont {Tong}\ and\ \citenamefont
  {Toshima}(2010{\natexlab{a}})}]{tong2010stru}%
  \BibitemOpen
  \bibfield  {author} {\bibinfo {author} {\bibfnamefont {X.~M.}\ \bibnamefont
  {Tong}}\ and\ \bibinfo {author} {\bibfnamefont {N.}~\bibnamefont {Toshima}},\
  }\href@noop {} {\bibfield  {journal} {\bibinfo  {journal} {Physical Review
  A}\ }\textbf {\bibinfo {volume} {81}},\ \bibinfo {pages} {063403} (\bibinfo
  {year} {2010}{\natexlab{a}})}\BibitemShut {NoStop}%
\bibitem [{\citenamefont {Johnsson}\ \emph {et~al.}(2007)\citenamefont
  {Johnsson}, \citenamefont {Mauritsson}, \citenamefont {Remetter},
  \citenamefont {L'Huillier},\ and\ \citenamefont {Schafer}}]{johnsson2007}%
  \BibitemOpen
  \bibfield  {author} {\bibinfo {author} {\bibfnamefont {P.}~\bibnamefont
  {Johnsson}}, \bibinfo {author} {\bibfnamefont {J.}~\bibnamefont
  {Mauritsson}}, \bibinfo {author} {\bibfnamefont {T.}~\bibnamefont
  {Remetter}}, \bibinfo {author} {\bibfnamefont {A.}~\bibnamefont
  {L'Huillier}}, \ and\ \bibinfo {author} {\bibfnamefont {K.~J.}\ \bibnamefont
  {Schafer}},\ }\href@noop {} {\bibfield  {journal} {\bibinfo  {journal} {Phys.
  Rev. Lett.}\ }\textbf {\bibinfo {volume} {99}},\ \bibinfo {pages} {233001}
  (\bibinfo {year} {2007})}\BibitemShut {NoStop}%
\bibitem [{\citenamefont {Ranitovic}\ \emph {et~al.}(2010)\citenamefont
  {Ranitovic} \emph {et~al.}}]{ranitovic2010}%
  \BibitemOpen
  \bibfield  {author} {\bibinfo {author} {\bibfnamefont {P.}~\bibnamefont
  {Ranitovic}} \emph {et~al.},\ }\href@noop {} {\bibfield  {journal} {\bibinfo
  {journal} {New J. Phys.}\ }\textbf {\bibinfo {volume} {12}},\ \bibinfo
  {pages} {013008} (\bibinfo {year} {2010})}\BibitemShut {NoStop}%
\bibitem [{\citenamefont {Tong}\ \emph {et~al.}(2010)\citenamefont {Tong},
  \citenamefont {Ranitovic}, \citenamefont {Cocke},\ and\ \citenamefont
  {Toshima}}]{tong2010mech}%
  \BibitemOpen
  \bibfield  {author} {\bibinfo {author} {\bibfnamefont {X.~M.}\ \bibnamefont
  {Tong}}, \bibinfo {author} {\bibfnamefont {P.}~\bibnamefont {Ranitovic}},
  \bibinfo {author} {\bibfnamefont {C.~L.}\ \bibnamefont {Cocke}}, \ and\
  \bibinfo {author} {\bibfnamefont {N.}~\bibnamefont {Toshima}},\ }\href@noop
  {} {\bibfield  {journal} {\bibinfo  {journal} {Physical Review A}\ }\textbf
  {\bibinfo {volume} {81}},\ \bibinfo {pages} {021404} (\bibinfo {year}
  {2010})}\BibitemShut {NoStop}%
\bibitem [{\citenamefont {Riviere}\ \emph {et~al.}(2009)\citenamefont
  {Riviere}, \citenamefont {Uhden}, \citenamefont {Saalmann},\ and\
  \citenamefont {Rost}}]{riviere2009}%
  \BibitemOpen
  \bibfield  {author} {\bibinfo {author} {\bibfnamefont {P.}~\bibnamefont
  {Riviere}}, \bibinfo {author} {\bibfnamefont {O.}~\bibnamefont {Uhden}},
  \bibinfo {author} {\bibfnamefont {U.}~\bibnamefont {Saalmann}}, \ and\
  \bibinfo {author} {\bibfnamefont {J.~M.}\ \bibnamefont {Rost}},\ }\href@noop
  {} {\bibfield  {journal} {\bibinfo  {journal} {New Journal of Physics}\
  }\textbf {\bibinfo {volume} {11}},\ \bibinfo {pages} {053011} (\bibinfo
  {year} {2009})}\BibitemShut {NoStop}%
\bibitem [{\citenamefont {Holler}\ \emph {et~al.}(2011)\citenamefont {Holler},
  \citenamefont {Schapper}, \citenamefont {Gallmann},\ and\ \citenamefont
  {Keller}}]{holler2011}%
  \BibitemOpen
  \bibfield  {author} {\bibinfo {author} {\bibfnamefont {M.}~\bibnamefont
  {Holler}}, \bibinfo {author} {\bibfnamefont {F.}~\bibnamefont {Schapper}},
  \bibinfo {author} {\bibfnamefont {L.}~\bibnamefont {Gallmann}}, \ and\
  \bibinfo {author} {\bibfnamefont {U.}~\bibnamefont {Keller}},\ }\href@noop {}
  {\bibfield  {journal} {\bibinfo  {journal} {Physical Review Letters}\
  }\textbf {\bibinfo {volume} {106}},\ \bibinfo {pages} {123601} (\bibinfo
  {year} {2011})}\BibitemShut {NoStop}%
\bibitem [{\citenamefont {Chu}\ and\ \citenamefont {Telnov}(2004)}]{chu2004}%
  \BibitemOpen
  \bibfield  {author} {\bibinfo {author} {\bibfnamefont {S.~I.}\ \bibnamefont
  {Chu}}\ and\ \bibinfo {author} {\bibfnamefont {D.~A.}\ \bibnamefont
  {Telnov}},\ }\href@noop {} {\bibfield  {journal} {\bibinfo  {journal} {Phys.
  Rep.}\ }\textbf {\bibinfo {volume} {390}},\ \bibinfo {pages} {1} (\bibinfo
  {year} {2004})}\BibitemShut {NoStop}%
\bibitem [{\citenamefont {Tong}\ and\ \citenamefont
  {Toshima}(2010{\natexlab{b}})}]{tong2010floq}%
  \BibitemOpen
  \bibfield  {author} {\bibinfo {author} {\bibfnamefont {X.~M.}\ \bibnamefont
  {Tong}}\ and\ \bibinfo {author} {\bibfnamefont {N.}~\bibnamefont {Toshima}},\
  }\href@noop {} {\bibfield  {journal} {\bibinfo  {journal} {Phys. Rev. A}\
  }\textbf {\bibinfo {volume} {81}},\ \bibinfo {pages} {043429} (\bibinfo
  {year} {2010}{\natexlab{b}})}\BibitemShut {NoStop}%
\bibitem [{\citenamefont {Ranitovic}\ \emph {et~al.}(2011)\citenamefont
  {Ranitovic}, \citenamefont {Tong}, \citenamefont {Hogle}, \citenamefont
  {Zhou}, \citenamefont {Liu}, \citenamefont {Toshima}, \citenamefont
  {Murnane},\ and\ \citenamefont {Kapteyn}}]{ranitovic2011}%
  \BibitemOpen
  \bibfield  {author} {\bibinfo {author} {\bibfnamefont {P.}~\bibnamefont
  {Ranitovic}}, \bibinfo {author} {\bibfnamefont {X.~M.}\ \bibnamefont {Tong}},
  \bibinfo {author} {\bibfnamefont {C.~W.}\ \bibnamefont {Hogle}}, \bibinfo
  {author} {\bibfnamefont {X.}~\bibnamefont {Zhou}}, \bibinfo {author}
  {\bibfnamefont {Y.}~\bibnamefont {Liu}}, \bibinfo {author} {\bibfnamefont
  {N.}~\bibnamefont {Toshima}}, \bibinfo {author} {\bibfnamefont {M.~M.}\
  \bibnamefont {Murnane}}, \ and\ \bibinfo {author} {\bibfnamefont {H.~C.}\
  \bibnamefont {Kapteyn}},\ }\href@noop {} {\bibfield  {journal} {\bibinfo
  {journal} {Physical Review Letters}\ }\textbf {\bibinfo {volume} {106}},\
  \bibinfo {pages} {193008} (\bibinfo {year} {2011})}\BibitemShut {NoStop}%
\end{thebibliography}%

% Optionally you can create manual references
%\begin{thebibliography}{19}
%\bibitem{lastie2005}	X. Y. Lastie, A. K. Grim, S. V. Moro, D. Jia, and A. A. Firs, {\it Nature} {\bf 400}, 197-200 (2005).
%\end{thebibliography}
% Produces the bibliography via BibTeX

\end{document}